%% file: main.tex
\documentclass[conference]{IEEEtran}

\usepackage{algorithm}
\usepackage{makecell}
\usepackage{url}
\usepackage{hyperref}
\usepackage{caption}
\usepackage{subcaption}
\usepackage{listings}
\usepackage{algpseudocode}
\usepackage{graphicx}
\usepackage{amsmath}
\usepackage{tikz}
\usetikzlibrary{arrows.meta,positioning,calc,fit} 

\newcommand{\anonymous}[1]{
    \ifdefined\isanonymoustrue
        \iffalse
            #1
        \else
            [Hidden]
        \fi
    \else
        #1
    \fi
}

\begin{document}

\title{Faster Distributed Inference-Only Recommender Systems via Bounded Lag Synchronous Collectives}
\author{
    \IEEEauthorblockN{Kiril Dichev\IEEEauthorrefmark{1}, Filip Igor Paw\l{}owski\IEEEauthorrefmark{1}, Albert-Jan Yzelman\IEEEauthorrefmark{1}}
    \IEEEauthorblockA{\IEEEauthorrefmark{1}Huawei Technologies Switzerland AG}
}

\maketitle

\begin{abstract}
Recommender systems are enablers of personalized content delivery, and therefore revenue, for many large companies. In the last decade, deep learning recommender models (DLRMs) are the de-facto standard in this field. The main bottleneck in DLRM inference is the lookup of sparse features across huge embedding tables, which are usually partitioned across the aggregate RAM of many nodes. In state-of-the-art recommender systems, the distributed lookup is implemented via irregular all-to-all (alltoallv) communication, and often presents the main bottleneck. Today, most related work sees this operation as a given; in addition, every collective is synchronous in nature. In this work, we propose a novel bounded lag synchronous (BLS) version of the alltoallv operation. The bound can be a parameter allowing slower processes to lag behind entire iterations before the fastest processes block. In special applications such as inference-only DLRM, the accuracy of the application is fully preserved. We implement BLS alltoallv in a new PyTorch Distributed backend and evaluate it with a BLS version of the reference DLRM code. We show that for well balanced, homogeneous-access DLRM runs our BLS technique does not offer notable advantages. But for unbalanced runs, e.g. runs with strongly irregular embedding table accesses or with delays across different processes, our BLS technique improves both the latency and throughput of inference-only DLRM. In the best-case scenario, the proposed reduced synchronisation can mask the delays across processes altogether.
\end{abstract}

\section{Introduction}

Recommender systems are the hidden infrastructure of the digital economy. They transform the information overload in the massive datasets into ranked, personalized choices at web scale. From apps, videos, song playlists, to product pages: recommender systems filter, rank, and curate content for each user while optimising multiple objectives such as satisfaction, retention and revenue. In practice, the ranking models are trained to predict engagement proxies such as click-through rate (CTR), i.e. the probability that a user clicks on a candidate item.

In recent years, Deep Learning Recommendation Models (DLRMs)~\cite{Naumov2019} have emerged as a state-of-the-art for high-dimensional sparse-feature tasks such as CTR prediction and feed ranking.
DLRM-style recommenders are key revenue drivers,
for the retail industry alone estimated at driving up to $5.7\%$ of revenue~\cite{chui18}. For online content companies, recommender systems -- albeit measured before the advent of pervasive LLM usage -- occupied up to $79\%$~\cite{gupta2020}.
DLRM systems process both continuous dense features (e.g., age, ratings) and categorical sparse features (e.g., user/item IDs, device type).
Dense features are transformed by a multilayer perceptron (MLP), while sparse features are mapped to low-dimensional dense vectors via large embedding tables. 
The resulting dense vectors are combined by an explicit dot-interaction layer, the output of which are fed to another MLP that produces the final prediction.
The former MLP is called the bottom or dense MLP, while the latter is known as the top or output MLP~\cite{Naumov2019}.
The growth in the number of parameters has repeatedly brought about improved accuracy~\cite{lian2021}.
The retrieval of embedding table entries (and update during training) is memory bound, whereas the remaining MLP computations account for most of the floating-point operations and is compute bound.
Since embedding tables can constitute over 99.99\% of the total parameters in production, DLRM workloads are memory-bandwidth bound. If embedding tables are distributed across nodes, the workload furthermore becomes communication bound.
For such cases, we propose a novel so called \textbf{bounded lag synchronous (BLS)} execution for DLRM inference, with the support of a corresponding BLS collective, and improve the robustness to either compute, memory, or communication imbalances.
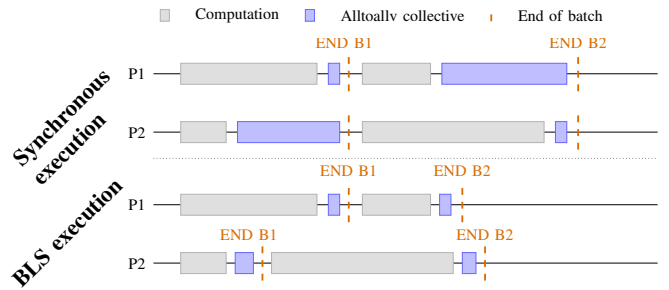
\begin{figure}[!htbp]
  \centering
  \resizebox{\columnwidth}{!}{\input{intro-bound.tikz}}
  \caption{Timeline comparison for two types of executions: Synchronous execution (top), and bounded lag synchronous execution (bottom). The BLS execution relies on a less synchronous alltoallv (up to a bound $k$) collective, allowing P1 and P2 stages to be decoupled.}
  \label{fig:alltoallv-compare}
\end{figure}

Consider the inference-only case of a distributed DLRM application (our reference is Naumov et al.~\cite{Naumov2019}).
The implementation employs a row-block partitioning when processing a batch of inputs, such that each rank is assigned a row-wise block partition of the batch, also called a \emph{mini-batch}.
The embedding tables are distributed by a given partitioning strategy.
For its minibatch, each process determines the vector IDs required from other processes, and retrieves them through an \emph{alltoallv} collective.

Figure~\ref{fig:alltoallv-compare} illustrates how allowing for a bounded lag across iterations in the inference stages across processes shortens the end-to-end time of execution.
In the illustration, even a bound $k = 2$ iterations significantly accelerates the completion of a parallel computation: in existing collectives, due to their synchronous nature, different delays across processes P1 and P2 accumulate.
In contrast, the bounded lag collectives can mask individual delays across processes P1 and P2, leading to a faster completion.
One disadvantage is that both P1 and P2 need to buffer additional inference data while progressing through iterations, so that stragglers can perform later lookups.
Since for DLRM different processes synchronise only in the collective operations, e.g. alltoallv, the key requirement is to provide collectives supporting BLS execution.

The  main software artifacts of this work are \textbf{(a)} a BLS alltoallv collective in a new PyTorch Distributed backend, and \textbf{(b)} a BLS-enabled version of the DLRM application.
We will show that for some settings, existing backends can also benefit from BLS-enabled DLRM.
The bound parameter can be generalized:
Let the bound be $k$.
We allow two processes to have a lag of up to $k$ iterations, that is, any 2 processes may be in iterations $i$ and $j$, as long as $\|j-i\|\le k$.
The bound implies a memory overhead, however, as a multiple of $k$ inference-related buffers are stored at each process (these overheads amount to hundreds of KBs in the evaluation).

Our research contributions are summarised as follows. We:
\begin{itemize}
        \item design a novel BLS alltoallv collective, and expose it via PyTorch Distributed as a compliant distributed backend (Sec.~\ref{sec:bls-backend}).
        \item minimally extend DLRM to support BLS mechanisms (Sec.~\ref{sec:exp:dlrm});
        \item design new synthetic and real benchmarks for DLRM, and detail the entire software and hardware setup (Sec.~\ref{sec:setup})
        \item evaluate the alltoallv primitives, as well as DLRM with and without our BLS extensions, and with and without our backend (Sec.~\ref{sec:results})
\end{itemize}

\section{Related Work}
Collectives were first successfully standardised within the MPI standard~\cite{GROPP1996789}.
The design and realisation of all standard collectives and their variants remain accurately described by Chan et al.~\cite{collective-comm}.

As outlined in the previous section, the main obstacle in making inference-only DLRM more robust to imbalances is inherent synchronisation points.
In the case of distributed DLRM, the points of synchronisation between processes are the collective alltoallv exchanges.
Similar observations are also central for another recent contribution, which focuses on HPC applications~\cite{afzal2023making};
since most HPC applications have stronger data dependencies than our use case, the authors are constrained to using existing MPI collectives, and instead explore different existing variants.
Early work on DLRMs already recognised that standard collectives fulfill the role of exchanging vectors from sharded embedding tables, and integrated the standard optimised HPC collectives with their DLRM inference stack~\cite{Naumov2019,kalamkar20}.

The more recent \emph{non-blocking collectives}~\cite{hoefler2007implementation,mpi30} were a major step forward for MPI in overlapping computation and collective communication; this overlap is locked within the same iteration; DLRM uses the same overlap technique.
In a sense, our design further extends this idea to entire iterations, which enforces us to use additional application data buffers, a complication non-blocking collectives do not require.
For distributed DLRM, two types of parallelism are common and widely used: data parallelism and model parallelism.
As a special consideration of the model parallelism, where embedding tables are distributed, and accesses are expensive, sharding and caching play important roles.
Our bounded lag synchronous collectives are entirely orthogonal and complementary to the efforts of finding good sharding or caching strategies.

Meta and Nvidia are currently maintaining two leading recommender systems projects, TorchRec \cite{torchrec} and Merlin \cite{merlin}.
TorchRec is a framework for different recommender systems with advanced capabilities (e.g. sharding strategies); it employs a single-level caching scheme that stores often-visited rows of the embedding table on the accelerators.
Merlin is an extensive framework for recommender systems; for distributed embedding lookups, they rely on NCCL (Nvidia Collective Communications Library).
Merlin also contains useful utilities, such as NVTabular, which we use in our evaluation.
In terms of caching strategies, Merlin's HugeCTR supports distributed inference via a hierarchical parameter server: a multi-level embedding store using a hierarchy of different memories. 
Overall, the placement and caching strategies in use by TorchRec and Merlin reduce the latency and bandwidth costs of embedding lookups and help exploit the power-law access skew.

Park et al.~\cite{park2024} consider the acceleration of DLRM systems through a CXL interconnect. Cold embedding vectors are stored in CXL-attached memory, while often-accessed ones are kept in local accelerator memory. A dynamic mechanism separates and rebalances the two classes of vectors online.
Chen et al.\ accelerate DLRM inference through processing-in-memory, addressing memory-boundedness within the memory technology directly~\cite{chen24}. Huang et al.\ similar to us consider the all-to-all collective, but propose to use smart switches as a caching mechanism~\cite{10.1145/3677333.3678267}. Similarly, Guo et al.\ use smart NICs for caching, but also offer extensions that reduce memory lookup overheads~\cite{guo23}.
Further optimisations to DLRM systems consider training, algorithmic optimisation, compression and further memory system based approaches~\cite{ren25,wang24,adnan24,ye23,ardestani22,pumma21}. Beyond the contributions already mentioned, we are not aware of others focusing on redesigning the communication layer for accelerating this particular use case. Many of the preceding optimisations such as caching, algorithmic optimisations, job-level scheduling, and alternative hardware design are orthogonal to communication layer enhancements; i.e., our contributions may be applied simultaneously, and independently from most other optimisations we here summarised.

Our low-level one-sided implementation of alltoallv relies on Lightweight Parallel Foundations (LPF), a minimal set of primitives for high-performance communication~\cite{suijlen2019lightweightparallelfoundationsmodelcompliant}. A set of twelve primitives allows the low-level expression of RDMA requests; when composed, they support all higher-level communication patterns of practical interest without sacrificing performance. The use of LPF primitives adheres to well-defined cost semantics rooted in the Bulk Synchronous Parallel (BSP) model~\cite{bsp}, while simultaneously allowing for its functional and cost semantics to be extended to support any BSP-like model.
Alpert et al.\ describe cBSP where a global communication pattern is parametrised in process-local send/receive counts, such that the determination of round completion can be made by querying the local status only (e.g., local network interface counters)~\cite{alpert97}. To minimize the overheads of a global synchronisation, we use a cBSP-enabled version of LPF; the MPI community has also proposed similar extensions to the one-sided MPI primitives ~\cite{belli2015notified,Sergent2019}.
Another contribution targeting reduced synchronicity is the Stale Synchronous Parallel (SSP) model ~\cite{ssp_eric_xing,petuum}; it modifies the parameter server to allow a degree of staleness.
In SSP, stale versions of the same model weights (up to a degree of staleness) may be re-used, while still maintaining the convergence, even during training.
In our BLS technique, entirely different model parameters (i.e. embedding vectors) are loaded at each inference step instead, and are always fully accurate. As a result, in contrast to SSP, our idea so far cannot be applied to a training phase.

In terms of interoperability with existing software, it is desirable to comply with the MPI standard, and to implement e.g. low-level hardware optimisations within the communication library (exposing e.g. MPI collectives for accelerators~\cite{chen23}).
Non-blocking collective calls from the MPI standard could potentially be used for the purposes of our work, with some extensions (such as passing the bound $k$ at initialization), since buffered, pre-registered data is used by our collectives.
Our BLS-enabled DLRM version (Sect.~\ref{sec:exp:dlrm}) implements one way to approach that.
We comply with PyTorch Distributed~\cite{Li2020} in this work, which provides a Python-native interface to communication mechanisms -- including collectives -- for which Torch may provide different backends: MPI for standard collectives, NCCL for GPU accelerators, et cetera. Our BLS collectives are implemented in a new PyTorch backend.

\section{BLS Backend for PyTorch}
\label{sec:bls-backend}

PyTorch Distributed allows for the definition of backends, which provide PyTorch with the capability of deploying over different systems, architectures, and fabrics. We first describe the MPI backend, the default backend for CPU-only distributed runs. We then highlight the differences with our novel BLS backend, and detail our design and implementation.

\subsection{The MPI backend}

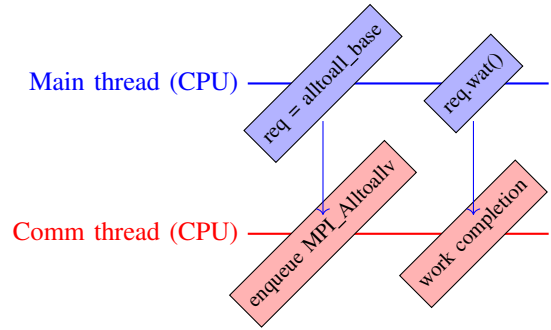
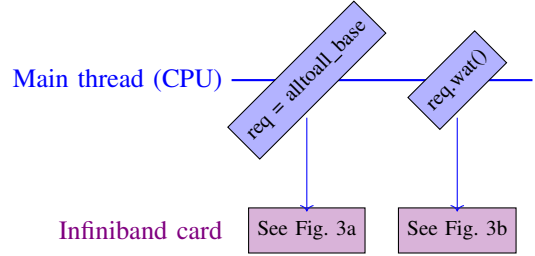
\begin{figure}
\centering
    \begin{subfigure}[b]{0.45\textwidth}
       \begin{tikzpicture}[
        event/.style={draw, rectangle, minimum size=0.6cm, fill=#1!30},
        thread/.style={thick, #1}
        ]
            \draw[thread=blue] (0,2.) -- (4,2.);
            \foreach \x/\text in {1/req = alltoall\_base, 3/req.wat()} {
                \node[event=blue,rotate=45,font=\footnotesize] at (\x,2.) {\text};
            }
            \draw[thread=red] (0,0) -- (4,0);
            \foreach \x/\text in {1/enqueue MPI\_Alltoallv, 3/work completion} {
                \node[event=red,rotate=45,font=\footnotesize] at (\x,0.) {\text};
                \draw[->,blue] (\x,1.5) -- (\x,0.25);
            }
            \node[blue, left] at (0,2.) {Main thread (CPU)};
            \node[red, left] at (0,0) {Comm thread (CPU)};
        \end{tikzpicture}
        \caption{The MPI backend of PyTorch Distributed uses a dedicated communication progress thread per process, which has performance caveats.}
        \label{fig:mpi_backend_async}
    \end{subfigure}
    \hspace{12pt}
    \begin{subfigure}[b]{0.45\textwidth}
             \begin{tikzpicture}[
        event/.style={draw, rectangle, minimum size=0.6cm, fill=#1!30},
        thread/.style={thick, #1}
        ]
            \draw[thread=blue] (0,2.) -- (4,2.);
            \foreach \x/\text in {1/req = alltoall\_base, 3/req.wat()} {
                \node[event=blue,rotate=45,font=\footnotesize] at (\x,2.) {\text};
            }
            
             \foreach \x/\text in {1/See Fig.~\ref{fig:init_and_complete:a}, 3/See Fig.~\ref{fig:init_and_complete:b}} {
                \node[event=violet,font=\footnotesize] at (\x,0.) {\text};
                \draw[->,blue] (\x,1.5) -- (\x,0.25);
            }
            \node[blue, left] at (0,2.) {Main thread (CPU)};
            \node[violet, left] at (0,0) {Infiniband card};
        \end{tikzpicture}  
        
        \caption{Our BLS backend for PyTorch Distributed offloads via non-blocking calls to the Infiniband card. No multi-threading is required.}
        \label{fig:lpf_backend_async}
    \end{subfigure}
    \caption{Illustration of how collectives are handled in the MPI and BLS backend in PyTorch Distributed. BLS offloads non-blocking RDMA calls to the NIC and reduces synchronicity.}
    \label{fig:versus}
\end{figure}

PyTorch Distributed exposes collectives that can be invoked in blocking or non-blocking mode.
For the MPI backend, the synchronous mode calls a blocking version of an MPI collective operation and waits until it returns.
The asynchronous mode offloads the call to the dedicated communication progress thread, but still invokes a blocking version of an MPI collective operation.
The mechanism is visualised in Fig.~\ref{fig:mpi_backend_async}.
PyTorch Distributed does not leave it to the MPI implementation to asynchronously progress the communication, instead managing the asynchronous progress with dedicated progress threads in the Torch backend implementation.
We have two different concerns with this design: \textbf{(a)} We have observed that the progress thread can be a bottleneck for enqueuing new communication tasks. We have also observed it collapse in performance when used in connection with the OpenMP-parallel threads employed by embedding bag operations. Others~\cite[Sect.7D]{kalamkar20} have also reported related performance issues. \textbf{(b)} 
By definition of MPI collectives, \textit{in MPI, no thread shall leave a collective operation before all participating processes have at least entered it}.
In Fig.~\ref{fig:mpi_backend_async}, even if one communication thread is detached from its main threads, it is synchronised with all communication threads across the MPI communicator.

\input{bls_backend}

\input{bls_alltoallv}

\subsection{Bound: the parameter determining the degree of lag}
\label{sec:bound}

\begin{figure}
    \centering
    \begin{tikzpicture}[
    timeline/.style={thick, -Stealth},
    message/.style={thick, -Stealth, red},
    event/.style={circle, draw=black, fill=white, inner sep=2pt},
    time label/.style={above, font=\small}
]

\coordinate (A-start) at (0,0);
\coordinate (A-end) at (6.5,0);
\coordinate (B-start) at (0,-1);
\coordinate (B-end) at (6.5,-1);
\coordinate (C-start) at (0,-2);
\coordinate (C-end) at (6.5,-2);

\draw[timeline] (A-start) -- (A-end) node[right] {P0};
\draw[timeline] (B-start) -- (B-end) node[right] {P1};
\draw[timeline] (C-start) -- (C-end) node[right] {P2};

\foreach \x \t in {0/0, 2/1, 4/2} {
    \node[above] at (\x+1.,0.5) {\t};
    \draw[thin, gray!50] (\x,0) -- (\x,-2);
}


\node[event,red] (A1_b) at (1.5,0) {};
\node[event,green] (A1_a) at (0.5,0) {};
\node[] (A1-above) at (1.5,0.25) {S(*,*)};
\node[] (A1-2above) at (0.5,0.25) {R(*,*)};

\node[event,red] (A2) at (2.75,-2) {};
\node[event,green] (A2_2) at (3.5,-2) {};
\node[] (A2-above) at (3.5,-1.75) {S(*,*)};
\node[] (A2-2above) at (2.25,-1.75) {R(*,*)};

\node[event,red] (B3) at (4.75,-1) {};
\node[] (B3-above) at (4.25,-0.75) {R(*,*)};
\node[event,green] (B3_2) at (5.5,-1) {};
\node[] (B3_2-above) at (5.5,-0.75) {S(*,*)};

\node[] at (4,1) {Iter};

\end{tikzpicture}
    \caption{Illustration of a parallel run with $3$ processes and bound $k = 3$ (i.e., also $3$ receive buffers per process). P1 is ahead of P0 and P2, but needs to block and wait until $R(*,0)$ is available to read. Color scheme: red = not all operations are complete for process / iteration. green: operations are complete for process / iteration.} 
    \label{fig:bound-factor}
\end{figure}

Next, we consider the bound parameter and its importance for the permitted lag between processes; we also touch on the need for a corresponding global buffer count. 
Fig.~\ref{fig:bound-factor} illustrates an example for bound factor $3$, which accordingly requires three receive buffers $R(*,0)$, $R(*,1)$ and $R(*,2)$ per process.
$R(*,i)$ corresponds to a receive buffer $i$ with the memory locations corresponding to all ($*$) processes. P0, P1, and P2 are in iterations 0, 2, and 1, respectively.
As already outlined in Sect.~\ref{sec:alltoallv-init}, the modulo operator determines which buffer to use for which input.

The bound parameter determines not only the number of receive buffers needed per process; it also determines the permitted lag between processes.
In Fig.~\ref{fig:bound-factor},
we consider the green buffers as \textit{complete} and the red ones as \textit{pending}.
In the illustration, P1 has completed sends for iterations 0-2, but all its receive buffers are pending completion since P0 and P2 have not yet completed the alltoallv corresponding to iterations 0 and 1, respectively. Recall that P1 does not need to be synchronised with other processes; it may execute the bottom MLP for each of those three iterations independently of whether communication has been received. However, after completing the bottom MLP for iteration 2, P1 now has to block because it requires the communication for the first iteration in order to progress with the top MLP of iteration 0, after which its place in the buffer would be relinquished and become reusable for iteration 3.

When P0 thus finishes sending data corresponding to the alltoallv of iteration 0, and when that data has arrived at P1 (via a blocking \textit{wait()} call in PyTorch Distributed at P1), then P1 immediately completes the top MLP and prediction of iteration 0, frees and repurposes the buffer for iteration 3. Similarly, when P0 and P2 complete the sends of iteration 1, then after receipt the buffer space for that iteration is relinquished and repurposed, and so on. Hence the bound parameter directly determines both the number of buffers per process and the degree of lag between processes in a parallel computation: the fastest and slowest process in BLS DLRM inference can be at most as many iterations apart as the bound parameter.

\subsubsection*{A zero-bound parameter}
For $k=0$, BLS collectives are semantically equivalent to synchronous ones and hence our alltoallv is semantically equivalent to MPI\_Alltoallv in MPI.
Our collective with $k=0$ hence also applies to DLRM training or any other standard collective-based workload. However, our implementation has one further difference with most commonly-used collective implementations: 
we employ one-sided RDMA (puts) to realise collective communications, and do not rely on two-sided (send/receive) communication. This design decision is driven by the BLS idea, as two-sided primitives would introduce too much synchronisation.

\section{BLS-Enabled DLRM}\label{sec:exp:dlrm}

The original DLRM implementation supports an overlap between compute and communicate. As shown in Listing~\ref{code:dlrm_orig}, each process may overlap the alltoallv exchange with the bottom MLP computation, after which the alltoallv will block to completion.

\begin{lstlisting}[frame=single,language=Python,caption=Original DLRM code (snippet from forward function) ,basicstyle=\ttfamily\footnotesize,label=code:dlrm_orig]
ly = self.apply_emb(lS_o, lS_i,
                    self.emb_l, self.v_W_l)
a2a_req = ext_dist.alltoall(ly, self.n_emb_per_rank)
x = self.apply_mlp(dense_x, self.bot_l)
ly = a2a_req.wait()
# compute CTR
z = self.interact_features(x,ly)
p = self.apply_mlp(z, self.top_l)
\end{lstlisting}

\begin{figure*}
    \centering
    \includegraphics[width=0.9\linewidth]{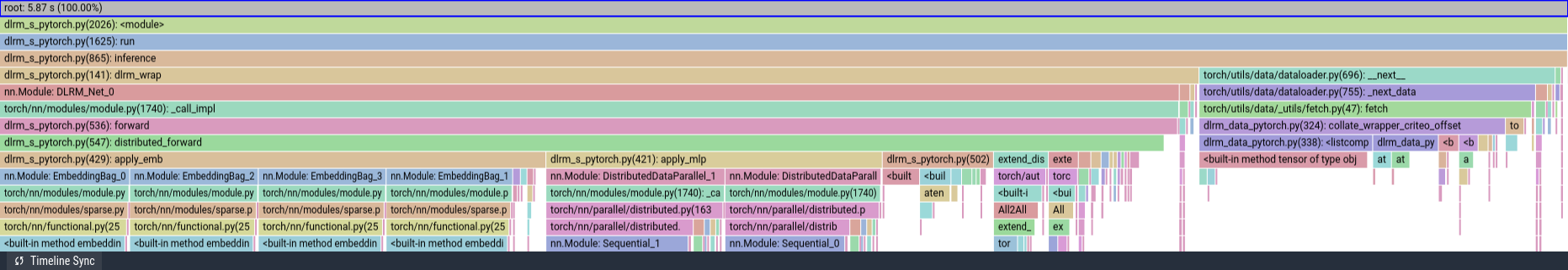}
    \caption{A flame graph showing a breakdown of the most time-consuming functions of an 8-process DLRM run with the original setup. Communication only takes $\approx 5 \%$ of execution time, but can become a bottleneck if execution is imbalanced.}
    \label{fig:flamegraph}
\end{figure*}

To validate that this snippet is representative for the main bottlenecks, we show a flamegraph of a standard DLRM run, generated via the PyTorch profiler, and rendered via Perfetto UI\footnote{\url{https://ui.perfetto.dev/}}, in Fig.~\ref{fig:flamegraph}.

We show a typical base run with our experimental setup, with eight nodes, one MPI process per node, reading process $0$ trace. 
The run takes $4.3$ seconds, and is dominated by embedding bag operations (apply\_emb), followed by apply\_mlp. Collective communication, the focus of our contribution, takes only $\approx 5$\% of execution time in this run. 
We disregard the expensive operation of fetching the next data item (stack on the right) and intentionally ignore it in further benchmark results, as it reflects on different issues, such as data generation or data I/O.

We release the original DLRM scheme from this restricted synchronicity and introduce a BLS DLRM version, as outlined in Listing~\ref{code:dlrm_mod}.
Our version allows two processes to run up to a bound $k$ iterations apart;
a process needs to block on the completion of the alltoallv operation if it had not already waited in the preceding $k$ iterations.
In case there are no stragglers delaying the alltoallv call completion, the call returns immediately. Otherwise the process blocks until the slowest straggler has completed the alltoallv from $k$ iterations ago (see Fig.~\ref{fig:bound-factor}).

\begin{lstlisting}[frame=single,language=Python,
caption=Our bounded lag synchronous DLRM code allowing non-blocking (via MPI backend) or BLS (via BLS backend) execution (snippet from forward function), basicstyle=\ttfamily\footnotesize,label=code:dlrm_mod]
ly = self.apply_emb(lS_o, lS_i,
                    self.emb_l, self.v_W_l)
# push to head of request list
a2a_req = ext_dist.alltoall(ly, self.n_emb_per_rank)
unfinished += 1
# x needs to be pushed into head of x-list
x = self.apply_mlp(dense_x, self.bot_l)
# Enforce wait for process in advanced iteration
# or it will overwrite buffer
if unfinished > bound:
    ly = a2a_req.wait() # wait on tail request
    <compute CTR>
    unfinished -= 1
# Drain all outstanding requests at the end
if last_batch:
    while unfinished > 0:
        ly = a2a_req.wait() # wait on tail request
        <compute CTR>
        unfinished -= 1
\end{lstlisting}

In addition to the shown snippets, the DLRM request class is extended to hold a FIFO queues of alltoallv requests as well as result tensors, instead of one request and one tensor. The wait function of the request class only returns the tensor at the tail, associated with the oldest alltoallv request.
In addition, we carefully remove any unnecessary collective calls, which would reintroduce synchronisation into our BLS design. 
For example, we eliminate a non-essential allgather operation used to provide diagnostic logging.

\begin{table}
    \centering
    \begin{tabular}{|p{3.cm}|p{1.cm}|p{1.5cm}|p{1.5cm}|}
    \hline
& \thead{DLRM\\ \\} & \thead{BLS DLRM\\with MPI\\backend} & \thead{BLS DLRM\\with BLS\\backend} 
    \\
    \hline
    overlaps compute and collective in same iteration? & Yes & Yes & Yes \\
    \hline
    overlaps compute and collective across iterations? & No & Yes & Yes \\
    \hline
    overlaps collective and collective across iterations? & No & No & Yes \\
    \hline
    \end{tabular}
    \caption{Gradually increasing degree of asynchronous support / overlap, first by extending DLRM, and then by introducing BLS backend.}
    \label{tab:dlrm_variants}
\end{table}

Our BLS DLRM version allows us to reduce the synchronicity of DLRM further than any existing version we are aware of. Different processes can overlap the progress of any stage across different iterations; with our BLS collective, this would also include overlapping collectives from different iterations.
This is illustrated in Table.~\ref{tab:dlrm_variants}.
The original DLRM code supports overlap of communicate and compute within the same iteration only.
Our contributions, BLS DLRM relying on either the MPI backend or the BLS backend, further increase the overlap capabilities, the maximum extent being via the BLS backend, where any two stages from two different iterations can overlap, even two communicate stages.

In summary, BLS has the potential to completely mask delays of different processes, as long as these delays can be masked within the bounded lag $k$.
If, however, the delays propagate beyond this point, a slowdown is unavoidable. 
For example, a single consistent straggler in a parallel run would lead to a poor performance of both synchronous and BLS execution. Another example where BLS would not be beneficial is a perfectly balanced application.

\section{Experimental Setup}
\label{sec:setup}

\subsection{Hardware}
\label{sec:exp:setup}

We use an eight-node ARM cluster for our experiments.
The nodes are ARM Kunpeng 920, with four NUMA domains spread over two sockets. 
Each domain has 24 cores, all clocked at 2.6 GHz; each domain has a $24$~MB L3 cache and $128$ of RAM, resulting in a total of $96$~MB L3 cache and $512$~GB RAM for the entire node.

All nodes are also equipped with \anonymous{a Huawei-manufactured network card, equipped with Mellanox (pre-Nvidia acquisition) ConnectX-5 chips}, and connected via 4x EDR Infiniband with $100$~Gbps maximum throughput. The eight nodes are connected with a single EDR switch.

We remark that our design idea does not require Infiniband in itself, but is relying on RDMA capabilities, such as the capability to poll for new messages in a message queue on hardware level.

\subsection{Software}
All nodes run Ubuntu 22.04.4 LTS. We use Spack for software management, and install PyTorch version 2.5.1 with MPI support. As MPI library, we use an open-source Open MPI-based distribution called \anonymous{ Hyper MPI, part of the HPC 22.0.0 package\cite{hyper-mpi}; it is compatible with Open MPI version 4.1.1rc4}.

For reproducibility and alignment with prior work (e.g., Gupta et al.~\cite{gupta2020}), we build on the reference open-source DLRM PyTorch implementation~\cite{Naumov2019} to demonstrate the improvements in latency and throughput when using extended DLRM with our backends. Our commits are on top of DLRM commit 64063a3. We try to keep DLRM code changes minimal, and outlined the key differences in Listing~\ref{code:dlrm_mod}.
For some performance insights, we reuse the existing profiling capabilities of DLRM (see Fig.~\ref{fig:flamegraph}), but for latency and throughput metrics we manually instrument the inference stage.

For all of our DLRM-based experiments, we use $1$ OpenMP thread (OpenMP-parallelism is used by DLRM in embedding bag operations).
It is important to restrict thread parallelism, since our reduced synchronicity can lead to an oversubscription of cores from different embedding table operations.
We also note that our setting is the highest performing one (increasing the thread count reduces performance in all the settings we tested).

\subsection{DLRM-Based Metrics: Latency and Throughput}

Our metrics are latency per batch (in seconds) and throughput (in batches per second).
We manually instrument the duration of each batch iteration at each MPI process.
Since we display metrics compiled over multiple repetitions, multiple processes, and many iterations, we show how we compile latency and throughput in all DLRM-based plots:
\begin{itemize}
    \item Latency: each data point shows the mean (with 95\% CI values also shown) over all latency times $\{ L(i,j)\} $ ,
    where $L(i,j)$ is the already precomputed mean latency of run $i$ and process $j$. 
\item Throughput: each data point shows the throughput as the mean (with 95\% CI values also shown) over all $T(i)$, 
where $T(i) = \sum_{j \in procs}{\frac{\# batches}{{L(i,j)}}}$. The sum is used as the data parallelism over the processes increases the throughput.
\end{itemize}

We use 8-node (1 MPI process per node) experiments, and 5 repeated runs per data point.

We also note that we preload the datasets before starting measurements.
This is particularly important for randomized benchmarks, where the generation of on-the-fly randomized data proved exceptionally slow.

\subsection{Synthetic alltoallv benchmarks}
We have prepared pure alltoallv benchmarks for both backends, the existing MPI backend and our BLS backend.
As outlined earlier, our implementation of the alltoallv collective differs from existing implementations in the heavy reliance on one-sided communication, both for initiation and completion at each process, which differs in various aspects from two-sided communication (e.g. no eager or rendezvous protocol is required).
On the other hand, MPI also relies on a linear exchange for the alltoallv operation, which makes the comparison plausible (MPI's alltoall implementation has a larger range of algorithms available).
We believe this benchmark is relevant in its own right.

\subsection{Synthetic DLRM benchmarks}
We have designed two types of synthetic benchmarks for DLRM, which demonstrate different key aspects of the backends.

\subsubsection{Setting 1: Heterogeneous message size benchmarks}
\label{sec:hetero-sizes}
Apart from generating random data on the fly, the random generator is capable of distributing a variable number of embedding vectors accessed per embedding table, from 1 to a configurable maximum; we set the maximum to $100$ for our experiments.
This increases the heterogeneity of message sizes per alltoallv operation.
In contrast, if reading from a dataset, DLRM reads exactly 1 vector per table for all dataset-based executions, resulting in near equal-sized alltoallv (some differences do manifest due to the variable table count per process).

\subsubsection{Setting 2: Random delays}
In this experiment, we introduce (uniformly) random delays from $0$ to a configurable maximum in each inference step; we use in all experiments a range of $0-0.01$ seconds (the mean equalling the observed inference latency of $0.005$ seconds).
In order to not overload the two different random settings, we combine this random delays setting with the Mini-Kaggle datasets.

Our delays powerfully model a slowdown of any nature, including CPU or memory bottlenecks; the most likely DLRM stages that could have a slowdown are (a) apply\_emb, (b) alltoall exchange, or (c) apply\_mlp.

\subsection{Real datasets with DLRM}

We use 2 types of real datasets for our experiments:

\textbf{(A)} We use the Criteo Display Advertising Challenge dataset (Mini-Kaggle)~\cite{criteo-display} that corresponds to a week of ad-impression logs with click/no-click labels for CTR prediction. Each example includes 13 continuous dense fields and 26 categorical sparse fields, resulting in 26 embedding tables.

\textbf{(B)} Since the terabyte Criteo dataset had a corrupt dataset at the time of writing\footnote{See issue here: \url{https://huggingface.co/datasets/criteo/CriteoClickLogs/discussions/6}}, 
we explored another popular dataset -- the Alibaba Click and Conversion Prediction dataset~\cite{ma2018entirespacemultitaskmodel}. The conversion of this dataset to a Criteo-like dataset is not trivial, but most preprocessing steps are available in online tutorials~\cite{conversion-via-nvtabular}, and rely on tools such as NVTabular~\cite{nvtabular2025}. 
Our final converted dataset consists of 23 tables with categorical features.

We remark that both the Mini-Kaggle and Ali-CCP datasets result in relatively small embedding tables. 
The largest Mini-Kaggle table has approx. 1 million entries, and the largest Ali-CCP table has approx. 2 million entries.
For sparse embedding size of $s = 64 $ (bytes), this results in a few hundreds of MBs per embedding table for Ali-CCP.
The main advantage of the Ali-CCP dataset is that it provides a longer trace of user interactions, enabling larger batch sizes and longer benchmarks.
We are unable to find open, production datasets with larger embedding tables.
We also remark that NVTabular, the tool supporting the conversion of Ali-CCP, performs averaging of many-vector accesses to produce one-vector-per-table accesses. This renders the capability of BLS to exploit heterogeneous accesses (as in synthetic experiment of Sect.~\ref{sec:hetero-sizes}) useless for both used real datasets.

It is also important to quantify the additional memory space requirements of BLS DLRM.
If the embedding size is $s$, the batch size is $b$, and the embedding tables count is $\|tables\|$, we see the following requirements: As shown in Listing~\ref{code:dlrm_mod}, additional buffers of space $O(s^2 + b)$ need to be stored per iteration and process for later completion.
In addition, each process needs a buffer of space $O(s * b * \|tables\|)$ for each alltoallv operation (note that the process count is irrelevant).
These space requirements result in extra space of $O(k * (s * b * \|tables\| + s^2 + b))$ for a bound $k$.
For a typical run from our experiments with $b=512$, $s=64$, and $\|tables\| = 26$, this results in additional $\approx 860$ KB per process for each single increment of $k$.
A positive property of our additional buffers is their independence from the embedding table sizes; the table sizes motivate the distributed execution in the first place.

\section{Results}
\label{sec:results}

\subsection{Alltoallv benchmarks}
\label{sec:exp:synthetic}

We first measure in isolation the performance of the BLS alltoallv collective with $k=0$. As discussed, BLS alltoallv is based on one-sided put operations and differs from most other implementations which are based on two-sided communication. BLS alltoallv also differs from the one-sided BLS collectives released in LPF~\cite{lpf2025}; our completion relies on minimal synchronisation overheads (as proposed by cBSP~\cite{alpert97}). Our benchmark revolves around calls to the MPI\_Alltoallv signature, which then dispatches to the standard MPI implementation or to BLS alltoallv. We use the available 8 nodes, with MPI 1 process per node.

\begin{figure}
\centering
    \begin{subfigure}[b]{0.5\textwidth}        
        \includegraphics[width=\textwidth]{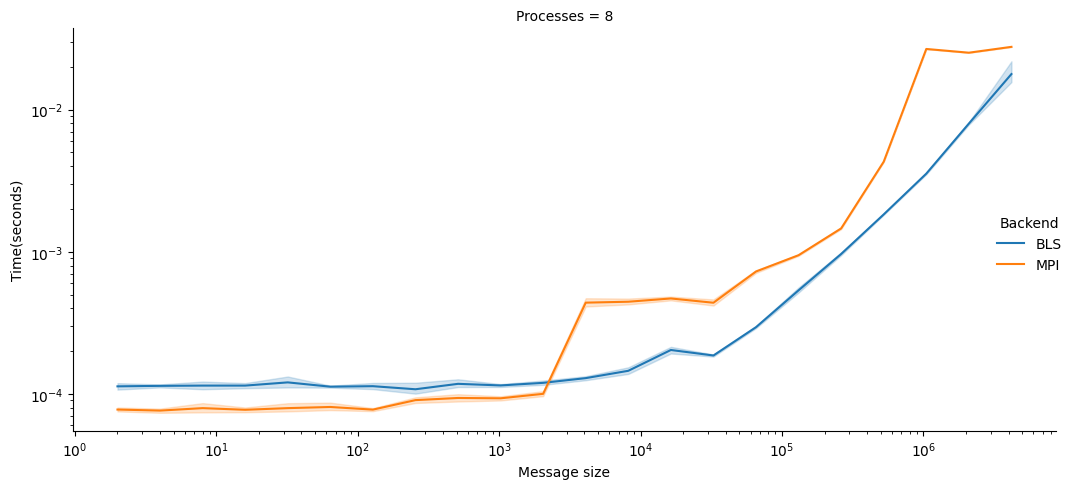}
        \caption{Time for varying message size per rank from 1B to $\approx 270$~MB (we use same-sized alltoallv in this case). Lower is better.}
        \label{fig:vary-sizes}
    \end{subfigure}
    \begin{subfigure}[b]{0.5\textwidth}
        \includegraphics[width=\textwidth]{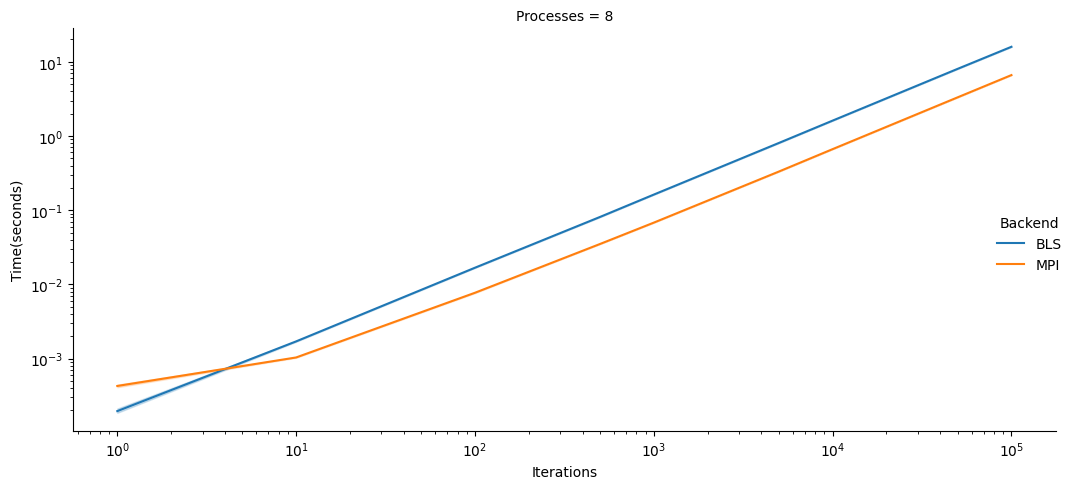}
        \caption{Time for varying repetitions of collective call, from 1 to 100K, with fixed 32KB-per-rank message size. Lower is better.}
        \label{fig:vary-iters}
    \end{subfigure}
    \caption{Comparison of our BLS alltoallv (bound $k = 0$) vs  MPI\_Alltoallv}
\end{figure}

In Fig.~\ref{fig:vary-sizes}, we compare the duration of one BLS alloallv vs one MPI\_Alltoallv, for message sizes per rank from 1B to $\approx 270$~MB (for simplicity, we rely on equal-sized alltoallv version semantically equivalent to alltoall).
MPI\_Alltoallv outperforms BLS alltoallv across small message sizes up to 4 KBs per rank, after which BLS alltoallv outperforms MPI\_Alltoallv across almost all message sizes (note the log-log scale).
In Fig.~\ref{fig:vary-iters}, we also show how BLS alltoallv performs compared to MPI\_Alltoallv when we use a randomly chosen fixed message size of $32$~kB per rank, but use a range of $1$ to $100$k collective calls in sequence.
BLS alltoallv only outperforms MPI\_Alltoallv for one-iteration benchmarks, but is outperformed by MPI\_Alltoallv with a factor of 2-3 for most runs.

We provide a speculative guess into the performance differences observed in either case.
Our implementation naively issues an IB work request for each RDMA put request, and additionally polls on the completion of each collective call.
MPI implementations may instead chain many small data transfers (including those to different target processes) into one IB work request. This may strongly contribute to the overheads of BLS alltoallv for small messages, where MPI performs better (Fig.~\ref{fig:vary-sizes}), as these overheads are relatively large to the data transmission.
On the other hand, there is a slowdown at $\approx 4$ KB per rank, where MPI loses in performance to BLS, and never recovers.
In two-sided communication, MPI performs a switch from eager to rendezvous protocol in that range; As the MPI collectives are implemented via two-sided communication, all MPI collectives are affected.
Our implementation instead relies on one-sided communication, where every exchange is immediate. This advantage remains for all larger messages.
Also, in Fig.~\ref{fig:vary-iters}, for a single iteration and one completion, the 32KB message alltoallv is more efficient for the BLS backend, similar to the same-size segment in Fig.~\ref{fig:vary-sizes}. However, with iterative repetitions of alltoallv calls, which consist of $p$ individual IB work requests at each rank, the cost of polling proportionally dominates by a constant factor, which is costly for many, frequent communication calls. In a future version, optimisations such as message combining (i.e. chaining IB requests) may lead to consistently faster BLS alltoallv across more settings.

Consistently outperforming the MPI variants is not the main objective of this work. Rather, we aim to implement a compatible collective, which shows benefits in its lower degree of synchronicity.
To eliminate some inefficiencies in comparison with MPI, a more efficient chaining of work requests and a less frequent polling on their completion could yield better performance.
For the BLS DLRM execution with the BLS backend, the suboptimal performance shown in Fig.~\ref{fig:vary-iters} is unlikely to manifest, as we do not issue collectives with this frequency; each collective is followed by operations on the dense and sparse neural network segments.
\subsection{DLRM benchmarks}
\subsubsection{Synthetic benchmarks: heterogeneous sizes and randomized delays}

\begin{figure}
\begin{subfigure}[b]{.5\textwidth}
\includegraphics[width=\textwidth]{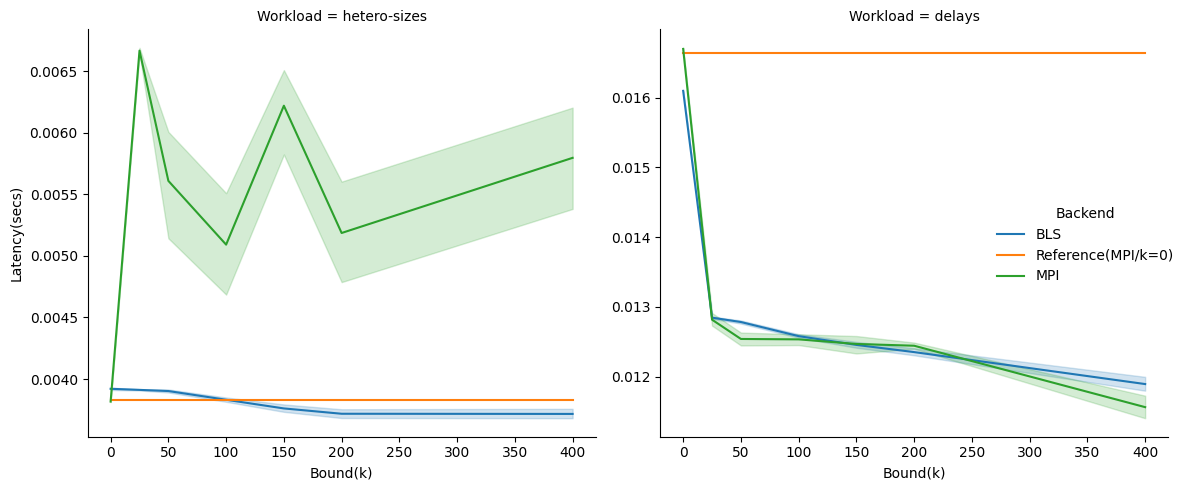}
\caption{Latency (lower is better)}
\end{subfigure}
\begin{subfigure}[b]{.5\textwidth}
\includegraphics[width=\textwidth]{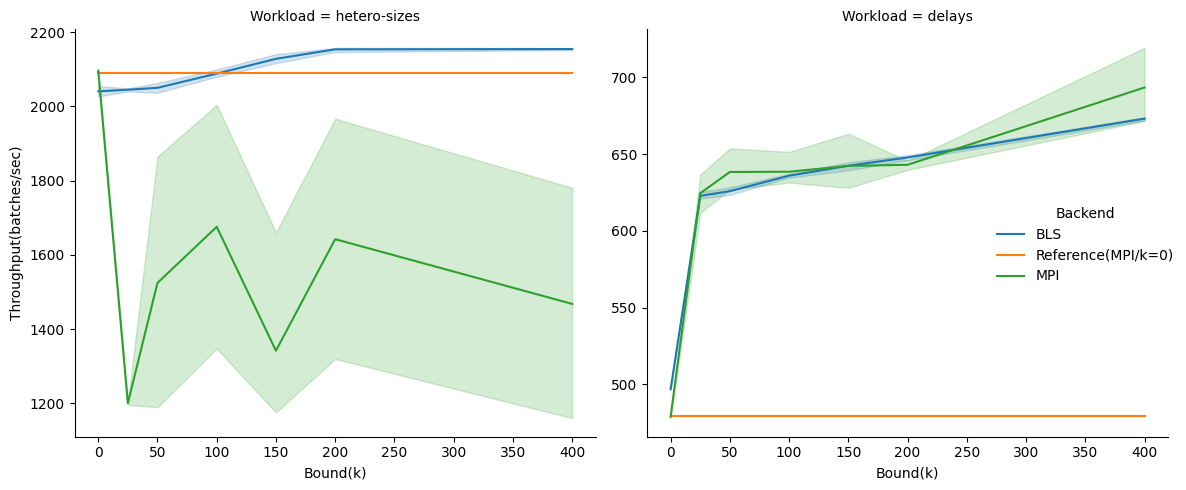}
\caption{Throughput (higher is better)}
\end{subfigure}
\caption{Synthetic benchmarks: Heterogeneous message sizes (random generation) and randomized delays. The increasing bound parameter has a positive impact on BLS backend for both benchmarks; for MPI backend, only delay-based benchmarks benefit.}
\label{fig:synth-bench}
\end{figure}

The results for both synthetic benchmarks are shown in Fig.~\ref{fig:synth-bench}.
For clarity, in the DLRM experiments we plot as a horizontal line the corresponding reference value for DLRM ($k = 0$) and the MPI backend; this is the reference we wish to outperform.
The heterogeneous message size benchmarks demonstrate that only the BLS backend benefits from increasing bound parameter, in contrast to the MPI backend. 
The end-to-end gains are small ($\approx 7\%$ lower latency and $\approx 6\%$ higher throughput), since communication does not really dominate execution, but they are reproducible.
We attribute this to two possible reasons.
On one hand, as we outlined in Tab.~\ref{tab:dlrm_variants}, only our backend can truly overlap different communications, which may be beneficial in strongly heterogeneous message exchanges.
On the other hand, we notice that the dedicated progress thread for the PyTorch MPI backend is slow in adding new communication requests; our backend is more efficient in this operation due to the immediate offload to the network card.
 We speculate that this contributes to the poor performance and increased variability of results with increasing bound parameter for the PyTorch MPI backend.

The second benchmark -- the randomized delays with homogeneous exchanges -- shows a very different behavior. 
In this case, both MPI and BLS backend benefit dramatically from a $k > 0$ bound parameter, with the gains quickly diminishing for larger bounds.
In this case, BLS DLRM benefits from both non-blocking MPI backend, or our BLS backend.
Note that the observed per-batch latency improves from $0.017$ to $0.012$, or by $0.005$, which is precisely the mean of the random distribution of delays per process and iteration.
In other words, the results demonstrate that our BLS-enabled DLRM \textit{completely masks} the delays across the execution for such cases; in this case, both backends (including the MPI backend) perform very well. The technique can therefore mask imbalances of various nature.
However, we note that if a single straggler consistently slows down execution, the delays cannot possibly be masked.

\subsubsection{Real dataset benchmarks}

\begin{figure}
\begin{subfigure}[b]{.5\textwidth}
\includegraphics[width=\textwidth]{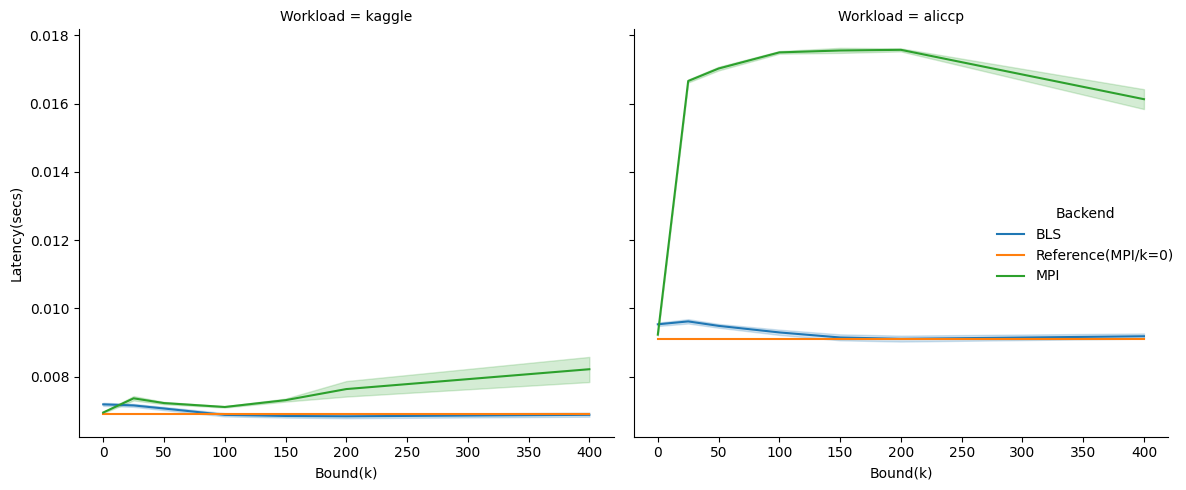}
\caption{Latency (lower is better)}
\end{subfigure}
\begin{subfigure}[b]{.5\textwidth}
\includegraphics[width=\textwidth]{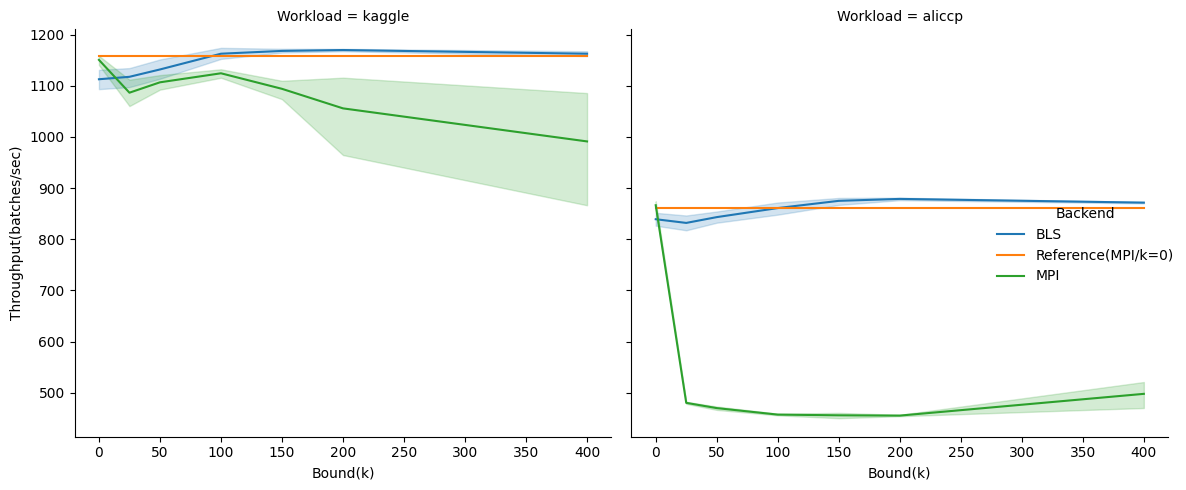}
\caption{Throughput (higher is better)}
\end{subfigure}
\caption{Real datasets: Mini-Kaggle and the converted Ali-CCP dataset. The BLS idea does not improve performance, possibly due to the well balanced execution for these datasets.}
\label{fig:real-bench}
\end{figure}

Our real benchmarks consist of two datasets: the popular Mini-Kaggle dataset and a manually converted Ali-CCP dataset.
The results are shown in Fig.~\ref{fig:real-bench}.
For both datasets, using a bound $ k = 0 $ leads to near-identical performance of either the MPI or the BLS backend.
Overall, for both Mini-Kaggle and Ali-CCP, no benefits can be seen from increasing the bound for either backend (see reference line).
This outcome is not surprising -- the DLRM datasets we explored run in a balanced manner, both in their compute/memory cost and in terms of their communication cost.
Therefore, while our BLS collectives improve performance wherever imbalances manifest, they do not seem beneficial for equally partitioned tables with equal compute/memory/communicate cost per node.

\section{Conclusion}

In this work, we designed a new bounded lag synchronous (BLS) alltoallv collective.
The BLS collective is synchronous for $k = 0$, with good performance for large messages; for $k > 0$ it allows for a lag between processes of entire iterations, within the bound $k$.
While not generally applicable in the latter case, it remains semantically correct in some important settings, such as inference-only DLRM.
We implemented this operation in a new lightweight BLS backend for PyTorch Distributed, which differs from the MPI backend in its truly asynchronous progress capabilities.
We take advantage of the BLS backend and its BLS collective in an extended version of DLRM, a BLS-enabled DLRM.
After performing various experiments, we find that $k > 0$ improves latency and throughput of inference \textbf{(a)} with randomly heterogeneous message exchanges for BLS DLRM + BLS alltoallv, and \textbf{(b)} with randomly occurring delays in execution for BLS DLRM and both BLS and MPI backend.
The improvements increase as non-uniform message sizes or uniformly distributed delays increase.
In contrast, for rather well balanced and homogeneous DLRM workloads, such as the tested Mini-Kaggle and Ali-CCP datasets, our BLS extensions do not consistently improve latency or throughput, compared to the original DLRM version with the MPI backend.

Overall, we believe that the BLS technique can transparently mask delays for distributed recommendation systems, improving both inference latency and inference throughput until they converge to those of fully balanced runs.
\section*{Acknowledgements}
      We acknowledge the use of DeepSeek (\url{https://deepseek.ai/}), which proposed the term Bounded Lag Synchronous after we outlined our design and implementation. It also was used to generate early versions of our Tikz/PGF plots.
We would like to thank our colleagues from \anonymous{the Huawei Z\"urich Research Center for the helpful discussions on DLRM and performance evaluation.}
\bibliography{refs}

\end{document}

%% file: intro-bound.tikz
\begin{tikzpicture}[x=0.35cm,y=0.9cm,font=\scriptsize]
\tikzset{
  lane/.style={line width=.4pt},
  comp/.style={fill=gray!25,draw=gray!60},
  commbox/.style={fill=blue!25,draw=blue!60},
  commline/.style={blue!65,very thick},
  endtick/.style={dashed,orange!85!black,thick}
}

\def\dividy{1.55}       
\def\panelpad{0.22cm}   
\def\labelpad{1.5pt}    

\draw (-0.2,3.0) -- (22,3.0); \node[anchor=east] at (-0.2,3.0) {P1};
\draw (-0.2,2.0) -- (22,2.0); \node[anchor=east] at (-0.2,2.0) {P2};
\node[anchor=east,font=\bfseries,rotate=45] at (-1.5,3.) {\parbox{2cm}{Synchronous execution}};

\draw[comp] (1.0,2.82) rectangle (7.0,3.18);       
\draw[commbox] (7.5,2.82) rectangle (8.0,3.18);       
\draw[comp] (1.0,1.82) rectangle (3.0,2.18);       
\draw[commbox] (3.5,1.82) rectangle (8.0,2.18);    
\draw[endtick] (8.4,2.70) -- (8.4,3.30);
\node[orange!85!black,above] at (8.2,3.30) {\scriptsize END B1};
\draw[endtick] (8.4,1.70) -- (8.4,2.30);

\draw[comp] (9.0,2.82) rectangle (12.0,3.18);      
\draw[commbox] (12.5,2.82) rectangle (18.0,3.18);  
\draw[comp] (9.0,1.82) rectangle (17.0,2.18);      
\draw[commbox] (17.5,1.82) rectangle (18, 2.18);
\draw[endtick] (18.5,2.70) -- (18.5,3.30);
\node[orange!85!black,above] at (18.5,3.30) {\scriptsize END B2};
\draw[endtick] (18.5,1.70) -- (18.5,2.30);

\draw[densely dotted,gray] (-0.2,\dividy) -- (22,\dividy);

\begin{scope}[yshift=-\panelpad]
  \draw (-0.2,1.0) -- (22,1.0); \node[anchor=east] at (-0.2,1.0) {P1};
  \draw (-0.2,0.0) -- (22,0.0); \node[anchor=east] at (-0.2,0.0) {P2};
  \node[anchor=east,font=\bfseries,rotate=45] at (-1.5,1.5) {BLS execution};

  \draw[comp]    (1.0,0.82) rectangle (7.0,1.18);
  \draw[commbox] (7.5,0.82) rectangle (8.,1.18);
  \draw[endtick] (8.4,0.70) -- (8.4,1.30);
  \node[orange!85!black,above=\labelpad] at (8.4,1.30) {\scriptsize END B1};
  \draw[comp]    (9.0,0.82) rectangle (12.0,1.18);
  \draw[commbox] (12.4,0.82) rectangle (12.9,1.18);
  \draw[endtick] (13.4,0.70) -- (13.4,1.30);
  \node[orange!85!black,above=\labelpad] at (13.4,1.30) {\scriptsize END B2};

  \draw[comp]    (1.0,-0.18) rectangle (3.0,0.18);
  \draw[commbox] (3.4,-0.18) rectangle (4.2,0.18);
  \draw[endtick] (4.6,-0.30) -- (4.6,0.30);
  \node[orange!85!black,above=\labelpad] at (4,0.20) {\scriptsize END B1};
  \draw[comp]    (5.0,-0.18) rectangle (13.0,0.18);
  \draw[commbox] (13.4,-0.18) rectangle (14.,0.18);
  \draw[endtick] (14.4,-0.30) -- (14.4,0.30);
  \node[orange!85!black,above=\labelpad] at (14.4,0.20) {\scriptsize END B2};
\end{scope}

\begin{scope}[shift={(10,4.0)}]
  \matrix [draw=none,fill=white,row sep=0pt,column sep=8pt,
           nodes={anchor=center,font=\scriptsize}] at (0,0) {
    \draw[comp] (0,0.1) rectangle (0.4,-0.1); & \node{Computation}; & 
    \draw[commbox] (0,0.1) rectangle (0.4,-0.1); & \node{Alltoallv collective}; & 
    \draw[endtick] (0.2,-0.1) -- (0.2,0.1); & \node{End of batch}; \\
  };
\end{scope}
\end{tikzpicture}

%% file: bls_backend.tex
\subsection{The BLS backend}
Fig.~\ref{fig:lpf_backend_async} illustrates how the main thread interacts with our BLS backend, which differs by design in two important aspects.
First, there is no need for a dedicated progress thread for communication, since RDMA calls are non-blocking and offload immediately from the CPU to the network card.
Second, all our collectives are bounded lag synchronous by design, which reduces inter-process synchronisation for bound $k > 0$.

Our backend relies on the MPI backend to provide common functionalities such as initialisation, and is not a full replacement of the MPI backend.
The BLS backend class hence contains an MPI backend member, to which we delegate some capabilities, such as MPI initialization or collective operations outside of alltoallv (e.g. barrier).
To provide BLS alltoallv, we implement the corresponding method in the BLS backend. Our implementation is tied to RDMA-capable hardware, such as Infiniband.

\subsubsection*{Initialisation}
Backend initialisation consists of the following steps:
\begin{itemize}
\item initialize MPI backend
\item create $k$ RDMA buffers (collective operation)
\item create $k$ tags
\end{itemize}

Before RDMA operations may occur, the remote memory for alltoallv transfers must be registered first. Our backend performs global registration only once, during initialisation; subsequent communication will re-use these buffers, relying on circular buffering.

Tags are low-level identifiers we use to annotate one-sided communication calls; they can be understood as \textit{colours} of every communication.
In this work, we use tags to ensure that different inference iterations do not interfere with each other.
We implement tags entirely within LPF~\cite{lpf2025}.
Tags are contained within the IB message headers with no additional overheads.
The tag-related counting introduces a small overhead; it is important that this mechanism is lightweight, since it is on the critical path of communication.

\subsubsection*{Initiating new alltoallv requests}
Every new alltoallv call starts a new BLS collective non-blockingly as described in Sect.~\ref{sec:bls-collective}, and returns a handle which also gets appended to a list of active requests. The call is compliant with the non-blocking call initiation for the MPI backend.

\subsubsection*{Completing alltoallv requests}
A wait() on each alltoallv handle is a blocking call, which completes the tail (or oldest) alltoallv request in the list, and pops it from the list. The completion of the alltoallv is detailed in Sect.~\ref{sec:bls-collective}. The call is compliant with the blocking completion call for the MPI backend.

%% file: bls_alltoallv.tex
\subsection{BLS Alltoallv}
\label{sec:bls-collective}

In DLRM training embedding tables get updated, which implies that during training processes need to be in sync in order to access the same updated weights.
The inference-only case for recommender systems is different.
Every inference iteration uses the lookups from a set of embedding vectors to compute the CTR predictions. 
The weights are already pretrained, and data is immutable. 
We are therefore released from the synchronicity requirements of general collectives.
We simply need to provide access to a multi-iteration set of required pretrained vectors.

Based on this information, we propose an alternative design for the distributed lookups, which we call  bounded lag synchronous (BLS) alltoallv.
The collective can operate on an extended set of buffers, enabling e.g. a process to offload one-sided puts across different-iteration alltoallv collectives without synchronising with its peers at all; no MPI implementation supports this.
That is, our BLS collective relaxes the barrier-style semantics of collectives, and tolerates a bounded lag $k$, allowing ranks to be up to $k$ iterations apart.

In order to be compliant with PyTorch Distributed employed by the reference DLRM implementation~\cite{Naumov2019} -- we employ the same two stages as the non-blocking collectives of the MPI backend:

\subsubsection*{Alltoallv initiation}
\label{sec:alltoallv-init}
\begin{figure}
    \begin{subfigure}[b]{0.45\textwidth}
    \centering
    \begin{tikzpicture}[
    node distance=2cm,
    recv-process/.style={draw, rectangle, minimum size=.3cm, thick, fill=blue!40},
    send-process/.style={draw, rectangle, minimum size=.3cm, thick, fill=blue!20},
    arrow/.style={-Stealth, thick, blue!60}
]

\coordinate (top-left) at (-1,1);
\coordinate (top-right) at (1,1);
\coordinate (bottom-left) at (-1,0);
\coordinate (bottom-right) at (1,0);

\node[send-process] (P1) at (top-left) {$S(0)$};
\node[recv-process] (P1_2) at (-1,2.2) {$R(0,i)$};
\node[recv-process] (P2) at (top-right) {$R(1,i)$};
\node[recv-process] (P3) at (bottom-left) {$R(3,i)$};
\node[recv-process] (P4) at (bottom-right) {$R(2,i)$};

\draw[arrow] (P1) -- (P2);
\draw[arrow] (P1) -- (P3);
\draw[arrow] (P1) -- (P4);
\draw[arrow] (P1) -- (P1_2);

\node[above, font=\small, blue!70] at ($(P1)!0.5!(P2)$) {tag $i$};
\node[left, font=\small, blue!70] at ($(P1)!0.5!(P3)$) {tag $i$};
\node[above, sloped, font=\small, blue!70] at ($(P1)!0.5!(P4)$) {tag $i$};
\node[right, font=\small, blue!70] at ($(P1)!0.5!(P1_2)$) {tag $i$};

\end{tikzpicture}
    \caption{Initiation of an alltoallv operation from the perspective of process 0.}
    \label{fig:init_and_complete:a}
    \end{subfigure}
    \hspace{12pt}
    \begin{subfigure}[b]{0.45\textwidth}
\centering
        \begin{tikzpicture}[
    node distance=2cm,
    recv-process/.style={draw, rectangle, minimum size=.3cm, thick, fill=blue!40},
    send-process/.style={draw, rectangle, minimum size=.3cm, thick, fill=blue!20},
    arrow/.style={-Stealth, thick, blue!60}
]

\coordinate (top-left) at (-1,1);
\coordinate (top-right) at (1,1);
\coordinate (bottom-left) at (-1,0);
\coordinate (bottom-right) at (1,0);

\node[recv-process] (P1) at (top-left) {$R(0,i)$};
\node[send-process] (P1_2) at (-1,2.2) {$S(0)$};
\node[send-process] (P2) at (top-right) {$S(1)$};
\node[send-process] (P3) at (bottom-left) {$S(3)$};
\node[send-process] (P4) at (bottom-right) {$S(2)$};

\draw[arrow] (P2) -- (P1);
\draw[arrow] (P3) -- (P1);
\draw[arrow] (P4) -- (P1);
\draw[arrow] (P1_2) -- (P1);

\node[above, font=\small, blue!70] at ($(P1)!0.5!(P2)$) {tag $i$};
\node[left, font=\small, blue!70] at ($(P1)!0.5!(P3)$) {tag $i$};
\node[above, sloped, font=\small, blue!70] at ($(P1)!0.5!(P4)$) {tag $i$};
\node[right, font=\small, blue!70] at ($(P1)!0.5!(P1_2)$) {tag $i$};

\end{tikzpicture}
    \caption{Completion of an alloallv operation from the perspective of process 0.}
        \label{fig:init_and_complete:b}
    \end{subfigure}
    \caption{$S(p)$ denotes local send buffer at process $p$. $R(p,i)$ denotes global receive buffer $i$ at process $p$. Tag $i$ denotes the tag ID of the message.}
\label{fig:init_and_complete}
\end{figure}

Fig.~\ref{fig:init_and_complete:a} illustrates, from the perspective of process 0, the required steps to initiate the alltoallv operation.
Each process follows the same logic, described as follows.
The process calls a linear sequence of puts into $comm\_size$ globally registered buffers. For each inference batch $j$, it targets receive buffers $R(p,i)$ with $i = (j \mod k)$ and $0 \leq p \leq (comm\_size-1)$. 
In addition, each put carries a tag $i$, an enabling mechanism for counting the messages per inference iteration later.
The put calls are non-blocking one-sided calls immediately offloaded to the network card, and require no additional threads.
No synchronisation happens between any two processes at any stage, a fundamental difference from how the MPI backend implements collectives as a sequence of two-sided operations.

\subsubsection*{Alltoallv completion}
\label{sec:alltoallv-compl}
Fig.~\ref{fig:init_and_complete:b} illustrates the completion of an initiated alltoallv call from the perspective of process 0. Each process follows the same logic, described as follows.
The completion of the alltoallv call in batch $j$ is equivalent to a blocking polling mechanism on the network interface, which returns after the process has received $(comm\_size-1)$ messages from its peers, all of them with tag $i$, where $i = (j \mod k)$, and at receive buffer $R(0,i)$. 
Upon completion, we copy the internal buffer $R(0,i)$ of batch $j$ (with $i = (j \mod k)$) to the PyTorch receive buffer, and flag the completion of the asynchronous work. Note that no race condition can occur as all processes will block for completion in linear order of the requests.

The two outlined operations are implemented within our custom low-level backend, without delegating any communication to MPI.

%% file: refs.bib
@article{afzal2023making,
  title={Making applications faster by asynchronous execution: Slowing down processes or relaxing {MPI} collectives},
  author={Afzal, Ayesha and Hager, Georg and Markidis, Stefano and Wellein, Gerhard},
  journal={Future Generation Computer Systems},
  volume={148},
  pages={472--487},
  year={2023},
  publisher={Elsevier}
}

@misc{hyper-mpi,
  title = {{Hyper MPI}},
  howpublished = {\url{https://support.huawei.com/enterprise/en/doc/EDOC1100292512/f5c6f116/hyper-mpi}},
  year = {2025}, 
  note = "[Online; accessed June-2025]"
}

@misc{chui18,
    author={M. Chui and J. Manyika and M. Miremadi and N. Henke and R. Chung and P. Nel and S. Malhotra},
    title={Notes from the {AI} frontier insights from hundreds of use cases},
    year={2018},
    notes={\url{https://www.mckinsey.com/~/media/mckinsey/featured%20insights/artificial%20intelligence/notes%20from%20the%20ai%20frontier%20applications%20and%20value%20of%20deep%20learning/notes-from-the-ai-frontier-insights-from-hundreds-of-use-cases-discussion-paper.pdf}}
}

@misc{conversion-via-nvtabular,
year=2025,
howpublished={\url{https://nvidia-merlin.github.io/models/stable/examples/03-Exploring-different-models.html}}
}

@manual{mpi30,
    author = "{Message Passing Interface Forum}",
    title  = "{MPI}: A Message-Passing Interface Standard Version 3.0",
    url    = "https://www.mpi-forum.org/docs/mpi-3.0/mpi30-report.pdf",
    year   = 2023
}

@inproceedings{hoefler2007implementation,
  title={Implementation and performance analysis of non-blocking collective operations for {MPI}},
  author={Hoefler, Torsten and Lumsdaine, Andrew and Rehm, Wolfgang},
  booktitle={Proceedings of the 2007 ACM/IEEE conference on Supercomputing},
  pages={1--10},
  year={2007}
}

@inproceedings{belli2015notified,
  title={Notified access: Extending remote memory access programming models for producer-consumer synchronization},
  author={Belli, Roberto and Hoefler, Torsten},
  booktitle={2015 IEEE International Parallel and Distributed Processing Symposium},
  pages={871--881},
  year={2015},
  organization={IEEE}
}

@inproceedings{guo23,
author = {Guo, Anqi and Hao, Yuchen and Wu, Chunshu and Haghi, Pouya and Pan, Zhenyu and Si, Min and Tao, Dingwen and Li, Ang and Herbordt, Martin and Geng, Tong},
title = {Software-Hardware Co-design of Heterogeneous SmartNIC System for Recommendation Models Inference and Training},
year = {2023},
isbn = {9798400700569},
publisher = {Association for Computing Machinery},
address = {New York, NY, USA},
url = {https://doi.org/10.1145/3577193.3593724},
doi = {10.1145/3577193.3593724},
booktitle = {Proceedings of the 37th ACM International Conference on Supercomputing},
pages = {336–347},
numpages = {12},
location = {Orlando, FL, USA},
series = {ICS '23}
}

@INPROCEEDINGS{kalamkar20,
  author={Kalamkar, Dhiraj and Georganas, Evangelos and Srinivasan, Sudarshan and Chen, Jianping and Shiryaev, Mikhail and Heinecke, Alexander},
  booktitle={SC20: International Conference for High Performance Computing, Networking, Storage and Analysis}, 
  title={Optimizing Deep Learning Recommender Systems Training on CPU Cluster Architectures}, 
  year={2020},
  volume={},
  number={},
  pages={1-15},
  doi={10.1109/SC41405.2020.00047}
}

@INPROCEEDINGS{pumma21,
  author={Pumma, Sarunya and Vishnu, Abhinav},
  booktitle={2021 IEEE/ACM Workshop on Machine Learning in High Performance Computing Environments (MLHPC)}, 
  title={Semantic-Aware Lossless Data Compression for Deep Learning Recommendation Model {(DLRM)}}, 
  year={2021},
  volume={},
  number={},
  pages={1-8},
  doi={10.1109/MLHPC54614.2021.00006}
}

@inproceedings{10.1145/3677333.3678267,
author = {Huang, Songjun and Li, Yihong and Chen, Liangkun and Zhang, Xiaoxi and Liu, Shuo and Duan, Jingpu and Wu, Wenfei and Chen, Xu},
title = {Alleviating All-to-All Communication for Deep Learning Recommendation Model Inference},
year = {2024},
isbn = {9798400718021},
publisher = {Association for Computing Machinery},
address = {New York, NY, USA},
url = {https://doi.org/10.1145/3677333.3678267},
doi = {10.1145/3677333.3678267},
booktitle = {Workshop Proceedings of the 53rd International Conference on Parallel Processing},
pages = {104–105},
numpages = {2},
location = {Gotland, Sweden},
series = {ICPP Workshops '24}
}

@INPROCEEDINGS{adnan24,
  author={Adnan, Muhammad and Maboud, Yassaman Ebrahimzadeh and Mahajan, Divya and Nair, Prashant J.},
  booktitle={2024 ACM/IEEE 51st Annual International Symposium on Computer Architecture (ISCA)}, 
  title={Heterogeneous Acceleration Pipeline for Recommendation System Training}, 
  year={2024},
  volume={},
  number={},
  pages={1063-1079},
  doi={10.1109/ISCA59077.2024.00081}
}

@inproceedings{chen24,
author = {Chen, Sitian and Tan, Haobin and Zhou, Amelie Chi and Li, Yusen and Balaji, Pavan},
title = {{UpDLRM}: Accelerating Personalized Recommendation using Real-World {PIM} Architecture},
year = {2024},
isbn = {9798400706011},
publisher = {Association for Computing Machinery},
address = {New York, NY, USA},
url = {https://doi.org/10.1145/3649329.3658266},
doi = {10.1145/3649329.3658266},
booktitle = {Proceedings of the 61st ACM/IEEE Design Automation Conference},
articleno = {211},
numpages = {6},
location = {San Francisco, CA, USA},
series = {DAC '24}
}

@article{alpert97,
  title={{cBSP}: Zero-cost synchronization in a modified {BSP} model},
  author={Alpert, Richard and Philbin, James},
  journal={NEC Research Institute, Princeton, NJ, USA, Tech. Rep},
  pages={97--054},
  year={1997}
}

@inproceedings{chen23,
  title={{MPI-xCCL}: A portable {MPI} library over collective communication libraries for various accelerators},
  author={Chen, Chen-Chun and Shafie Khorassani, Kawthar and Kousha, Pouya and Zhou, Qinghua and Yao, Jinghan and Subramoni, Hari and Panda, Dhabaleswar K},
  booktitle={Proceedings of the SC'23 Workshops of the International Conference on High Performance Computing, Network, Storage, and Analysis},
  pages={847--854},
  year={2023}
}

@article{Li2020,
author = {Li, Shen and Zhao, Yanli and Varma, Rohan and Salpekar, Omkar and Noordhuis, Pieter and Li, Teng and Paszke, Adam and Smith, Jeff and Vaughan, Brian and Damania, Pritam and Chintala, Soumith},
title = {PyTorch distributed: experiences on accelerating data parallel training},
year = {2020},
issue_date = {August 2020},
publisher = {VLDB Endowment},
volume = {13},
number = {12},
issn = {2150-8097},
url = {https://doi.org/10.14778/3415478.3415530},
doi = {10.14778/3415478.3415530},
journal = {Proc. VLDB Endow.},
month = aug,
pages = {3005–3018},
numpages = {14}
}

@inproceedings{ye23,
author = {Ye, Haojie and Vedula, Sanketh and Chen, Yuhan and Yang, Yichen and Bronstein, Alex and Dreslinski, Ronald and Mudge, Trevor and Talati, Nishil},
title = {GRACE: A Scalable Graph-Based Approach to Accelerating Recommendation Model Inference},
year = {2023},
isbn = {9781450399180},
publisher = {Association for Computing Machinery},
address = {New York, NY, USA},
url = {https://doi.org/10.1145/3582016.3582029},
doi = {10.1145/3582016.3582029},
booktitle = {Proceedings of the 28th ACM International Conference on Architectural Support for Programming Languages and Operating Systems, Volume 3},
pages = {282–301},
numpages = {20},
location = {Vancouver, BC, Canada},
series = {ASPLOS 2023}
}

@INPROCEEDINGS{ardestani22,
  author={Ardestani, Ehsan K. and Kim, Changkyu and Lee, Seung Jae and Pan, Luoshang and Axboe, Jens and Rampersad, Valmiki and Agrawal, Banit and Yu, Fuxun and Yu, Ansha and Le, Trung and Yuen, Hector and Mudigere, Dheevatsa and Juluri, Shishir and Nanda, Akshat and Wodekar, Manoj and Nair, Krishnakumar and Naumov, Maxim and Petersen, Chris and Smelyanskiy, Mikhail and Rao, Vijay},
  booktitle={2022 IEEE 42nd International Conference on Distributed Computing Systems (ICDCS)}, 
  title={Supporting Massive {DLRM} Inference through Software Defined Memory}, 
  year={2022},
  volume={},
  number={},
  pages={302-312},
  doi={10.1109/ICDCS54860.2022.00037}
}

@INPROCEEDINGS{ren25,
  author={Ren, Jie and Ma, Bin and Yang, Shuangyan and Francis, Benjamin and Ardestani, Ehsan K. and Si, Min and Li, Dong},
  booktitle={2025 IEEE International Symposium on High Performance Computer Architecture (HPCA)}, 
  title={Machine Learning-Guided Memory Optimization for {DLRM} Inference on Tiered Memory}, 
  year={2025},
  volume={},
  number={},
  pages={1631-1647},
  doi={10.1109/HPCA61900.2025.00121}
}

@INPROCEEDINGS{wang24,
  author={Wang, Weihu and Xia, Yaqi and Yang, Donglin and Zhou, Xiaobo and Cheng, Dazhao},
  booktitle={SC24: International Conference for High Performance Computing, Networking, Storage and Analysis}, 
  title={Accelerating Distributed {DLRM} Training with Optimized TT Decomposition and Micro-Batching}, 
  year={2024},
  volume={},
  number={},
  pages={1-15},
  doi={10.1109/SC41406.2024.00055}
}

@inproceedings{park2024,
  title={Accelerating Large-Scale DLRM Inference through Dynamic Hot Data Rearrangement},
  author={Park, Taehyung and Yang, Seungjin and Seok, Jongmin and Lee, Hyuk-Jae and Kim, Juhyun and Rhee, Chae Eun},
  booktitle={2024 IEEE International Symposium on Circuits and Systems (ISCAS)},
  pages={1--5},
  year={2024},
  organization={IEEE}
}

@article{Naumov2019,
  title={Deep learning recommendation model for personalization and recommendation systems},
  author={Naumov, Maxim and Mudigere, Dheevatsa and Shi, Hao-Jun Michael and Huang, Jianyu and Sundaraman, Narayanan and Park, Jongsoo and Wang, Xiaodong and Gupta, Udit and Wu, Carole-Jean and Azzolini, Alisson G and others},
  journal={arXiv preprint arXiv:1906.00091},
  year={2019}
}

@article{lian2021,
  title={Persia: a hybrid system scaling deep learning based recommenders up to 100 trillion parameters},
  author={Lian, Xiangru and Yuan, Binhang and Zhu, Xuefeng and Wang, Yulong and He, Yongjun and Wu, Honghuan and Sun, Lei and Lyu, Haodong and Liu, Chengjun and Dong, Xing and others},
  journal={arXiv preprint arXiv:2111.05897},
  year={2021}
}

@misc{criteo-display,
    author = {Jean-Baptiste Tien and joycenv and Olivier Chapelle},
    title = {Display Advertising Challenge},
    year = {2014},
    howpublished = {\url{https://kaggle.com/competitions/criteo-display-ad-challenge}},
    note = {Kaggle}
}

@inproceedings{merlin,
author = {Wang, Zehuan and Wei, Yingcan and Lee, Minseok and Langer, Matthias and Yu, Fan and Liu, Jie and Liu, Shijie and Abel, Daniel G. and Guo, Xu and Dong, Jianbing and Shi, Ji and Li, Kunlun},
title = {Merlin {HugeCTR}: {GPU}-accelerated Recommender System Training and Inference},
year = {2022},
isbn = {9781450392785},
publisher = {Association for Computing Machinery},
address = {New York, NY, USA},
url = {https://doi.org/10.1145/3523227.3547405},
doi = {10.1145/3523227.3547405},
booktitle = {Proceedings of the 16th ACM Conference on Recommender Systems},
pages = {534–537},
numpages = {4},
location = {Seattle, WA, USA},
series = {RecSys '22}
}

@inproceedings{torchrec,
author = {Ivchenko, Dmytro and Van Der Staay, Dennis and Taylor, Colin and Liu, Xing and Feng, Will and Kindi, Rahul and Sudarshan, Anirudh and Sefati, Shahin},
title = {{TorchRec}: a {PyTorch} Domain Library for Recommendation Systems},
year = {2022},
isbn = {9781450392785},
publisher = {Association for Computing Machinery},
address = {New York, NY, USA},
url = {https://doi.org/10.1145/3523227.3547387},
doi = {10.1145/3523227.3547387},
booktitle = {Proceedings of the 16th ACM Conference on Recommender Systems},
pages = {482–483},
numpages = {2},
location = {Seattle, WA, USA},
series = {RecSys '22}
}

@inproceedings{ssp_eric_xing,
 author = {Ho, Qirong and Cipar, James and Cui, Henggang and Lee, Seunghak and Kim, Jin Kyu and Gibbons, Phillip B. and Gibson, Garth A and Ganger, Greg and Xing, Eric P},
 booktitle = {Advances in Neural Information Processing Systems},
 editor = {C.J. Burges and L. Bottou and M. Welling and Z. Ghahramani and K.Q. Weinberger},
 pages = {},
 publisher = {Curran Associates, Inc.},
 title = {More Effective Distributed ML via a Stale Synchronous Parallel Parameter Server},
 url = {https://proceedings.neurips.cc/paper_files/paper/2013/file/b7bb35b9c6ca2aee2df08cf09d7016c2-Paper.pdf},
 volume = {26},
 year = {2013}
}

@misc{lpf2025,
title = {LPF: Lightweight Parallel Foundations},
year = 2025,
howpublished = {\url{https://github.com/Algebraic-Programming/LPF/}},
}

@misc{nvtabular2025,
year = 2025,
title = {NVTabular},
howpublished = {\url{https://github.com/NVIDIA-Merlin/NVTabular}}
}

@misc{ma2018entirespacemultitaskmodel,
      title={Entire Space Multi-Task Model: An Effective Approach for Estimating Post-Click Conversion Rate}, 
      author={Xiao Ma and Liqin Zhao and Guan Huang and Zhi Wang and Zelin Hu and Xiaoqiang Zhu and Kun Gai},
      year={2018},
      eprint={1804.07931},
      archivePrefix={arXiv},
      primaryClass={stat.ML},
      url={https://arxiv.org/abs/1804.07931}, 
}

@article{GROPP1996789,
title = {A high-performance, portable implementation of the MPI message passing interface standard},
journal = {Parallel Computing},
volume = {22},
number = {6},
pages = {789-828},
year = {1996},
issn = {0167-8191},
doi = {https://doi.org/10.1016/0167-8191(96)00024-5},
url = {https://www.sciencedirect.com/science/article/pii/0167819196000245},
author = {William Gropp and Ewing Lusk and Nathan Doss and Anthony Skjellum}
}

@article{collective-comm,
author = {Chan, Ernie and Heimlich, Marcel and Purkayastha, Avi and van de Geijn, Robert},
title = {Collective communication: theory, practice, and experience: Research Articles},
year = {2007},
issue_date = {September 2007},
publisher = {John Wiley and Sons Ltd.},
address = {GBR},
volume = {19},
number = {13},
issn = {1532-0626},
journal = {Concurrency and Computation: Practice and Experience},
month = sep,
pages = {1749–1783},
numpages = {35}
}

@misc{suijlen2019lightweightparallelfoundationsmodelcompliant,
      title={Lightweight Parallel Foundations: a model-compliant communication layer}, 
      author={Wijnand Suijlen and A. N. Yzelman},
      year={2019},
      eprint={1906.03196},
      archivePrefix={arXiv},
      primaryClass={cs.DC},
      url={https://arxiv.org/abs/1906.03196}, 
}

@article{bsp,
author = {Valiant, Leslie G.},
title = {A bridging model for parallel computation},
year = {1990},
issue_date = {Aug. 1990},
publisher = {Association for Computing Machinery},
address = {New York, NY, USA},
volume = {33},
number = {8},
issn = {0001-0782},
url = {https://doi.org/10.1145/79173.79181},
doi = {10.1145/79173.79181},
journal = {Commun. ACM},
month = aug,
pages = {103–111},
numpages = {9}
}

@inproceedings{petuum,
author = {Xing, Eric P. and Ho, Qirong and Dai, Wei and Kim, Jin-Kyu and Wei, Jinliang and Lee, Seunghak and Zheng, Xun and Xie, Pengtao and Kumar, Abhimanu and Yu, Yaoliang},
title = {Petuum: A New Platform for Distributed Machine Learning on Big Data},
year = {2015},
isbn = {9781450336642},
publisher = {Association for Computing Machinery},
address = {New York, NY, USA},
url = {https://doi.org/10.1145/2783258.2783323},
doi = {10.1145/2783258.2783323},
booktitle = {Proceedings of the 21th ACM SIGKDD International Conference on Knowledge Discovery and Data Mining},
pages = {1335–1344},
numpages = {10},
location = {Sydney, NSW, Australia},
series = {KDD '15}
}

@INPROCEEDINGS{gupta2020,
  author={Gupta, Udit and Wu, Carole-Jean and Wang, Xiaodong and Naumov, Maxim and Reagen, Brandon and Brooks, David and Cottel, Bradford and Hazelwood, Kim and Hempstead, Mark and Jia, Bill and Lee, Hsien-Hsin S. and Malevich, Andrey and Mudigere, Dheevatsa and Smelyanskiy, Mikhail and Xiong, Liang and Zhang, Xuan},
  booktitle={2020 IEEE International Symposium on High Performance Computer Architecture (HPCA)}, 
  title={The Architectural Implications of Facebook's DNN-Based Personalized Recommendation}, 
  year={2020},
  volume={},
  number={},
  pages={488-501},
  doi={10.1109/HPCA47549.2020.00047}}

@inproceedings{Sergent2019,
author = {Sergent, Marc and Aitkaci, C\'{e}lia Tassadit and Lemarinier, Pierre and Papaur\'{e}, Guillaume},
title = {Efficient notifications for MPI one-sided applications},
year = {2019},
isbn = {9781450371759},
publisher = {Association for Computing Machinery},
address = {New York, NY, USA},
url = {https://doi.org/10.1145/3343211.3343216},
doi = {10.1145/3343211.3343216},
booktitle = {Proceedings of the 26th European MPI Users' Group Meeting},
articleno = {5},
numpages = {10},
location = {Z\"{u}rich, Switzerland},
series = {EuroMPI '19}
}
